\documentclass[twocolumn, aps, nofootinbib, prd, floats, floatfix, amsmath, amssymb, superscriptaddress, preprintnumbers,longbibliography]{revtex4-2} %

\usepackage{graphicx}  
\usepackage{dcolumn}   
\usepackage{chemarrow}
\usepackage{color}
\usepackage[dvipsnames]{xcolor}
\usepackage{mathrsfs}
\usepackage{float}
\usepackage{physics}
\usepackage{slashed}
\usepackage{latexsym}
\usepackage{comment}

\usepackage[colorlinks=true, linkcolor=blue, citecolor=purple]{hyperref}
\usepackage{ulem}
\usepackage{orcidlink}
\usepackage{appendix}

\usepackage{tabularx} 
\newcolumntype{C}[1]{>{\centering\let\newline\\\arraybackslash\hspace{0pt}}m{#1}}

\usepackage{hyperref}

\usepackage{fancyvrb}

\newcommand{\be}{\begin{equation}}
\newcommand{\ee}{\end{equation}}
\newcommand{\ba}{\begin{eqnarray}}
\newcommand{\ea}{\end{eqnarray}}

\newcommand{\gsim}{\mathrel{\hbox{\rlap{\lower.55ex \hbox {$\sim$}}
			\kern-.3em \raise.4ex \hbox{$>$}}}}
\newcommand{\lsim}{\mathrel{\hbox{\rlap{\lower.55ex \hbox {$\sim$}}
			\kern-.3em \raise.4ex \hbox{$<$}}}}

\hypersetup{colorlinks=true,
	breaklinks=true,
	pdfstartview=Fit,
	linkcolor=blue,
	citecolor=blue,
	urlcolor=blue}

\begin{document}

\title{High-Energy Nuclear Recoils from $2\to3$ Dark Matter Scattering}

\author{Rundong Fang}
\email{rundongfang@pku.edu.cn}
\affiliation{School of Physics and State Key Laboratory of Nuclear Physics and Technology, Peking University, Beijing 100871, China}
\affiliation{Center for High Energy Physics, Peking University,
Beijing 100871, China}

\author{Ji-Heng Guo}
\email{guojiheng@buaa.edu.cn}
\affiliation{School of Physics, Beihang University, Beijing 102206, China}

\author{Ming Li \orcidlink{0009-0007-7322-010X}}
\email{by2419116@buaa.edu.cn}
\affiliation{School of Physics, Beihang University, Beijing 102206, China}

\author{Jia Liu \orcidlink{0000-0001-7386-0253}}
\email{{jialiu@pku.edu.cn}}
\affiliation{School of Physics and State Key Laboratory of Nuclear Physics and Technology,
Peking University, Beijing 100871, China}
\affiliation{Center for High Energy Physics, Peking University,
Beijing 100871, China}

\author{Xiao-Ping Wang \orcidlink{0000-0002-2258-7741}}
\email{hcwangxiaoping@buaa.edu.cn}
\affiliation{School of Physics, Beihang University, Beijing 102206, China}

\begin{abstract}
Dark matter can scatter endothermically while remaining in its ground
state if the collision produces an additional massive particle.
Motivated by the high-energy nuclear-recoil candidate reported
by LUX-ZEPLIN, we study radiative $2\to3$ scattering,
$\chi N\to\chi N s$.
For heavy dark matter scattering on xenon,
producing a particle $s$ of mass $200$--$350~\mathrm{keV}$ requires
speeds in the high-velocity tail of the halo distribution and
suppresses low-energy recoils.
When inelastic $2\to2$ scattering takes dark matter to a state heavier by $m_s$, the two processes have identical recoil endpoints at each incident speed.
For slowly varying amplitudes, three-body phase space enhances the
 $2\to3$ spectrum at intermediate recoil energies and
suppresses it near both endpoints relative to the normalized inelastic $2\to2$ spectrum.
We illustrate this mechanism in a model where $s$ is a light scalar.
Using the LZ exposure and detection efficiency, we fix the coupling
product by requiring one expected signal event in the recoil-energy
window around the candidate. Our benchmark then predicts signal events in the recoil-energy bins below and above this window.
\end{abstract}

\maketitle

\section{Introduction}

The LUX-ZEPLIN (LZ) Collaboration reported a nuclear-recoil candidate
at $E_R=248\pm23_{\rm stat}\pm23_{\rm sys}~\mathrm{keV}$ in a region
of low expected background~\cite{LZ:2026axp}.
A widely studied interpretation invokes endothermic $2\to2$ scattering,
in which halo dark matter upscatters to a heavier state and the mass
splitting $\delta$ kinematically suppresses low-energy
recoils~\cite{Tucker-Smith:2001myb, Tucker-Smith:2004mxa}.
Early studies of the LZ candidate explored this kinematic mechanism
and its realization in Higgsino dark
matter~\cite{Su:2026rwz,Fan:2026kxx,Freese:2026sga,DiMauro:2026ldr, Pospelov:2026ewn, Wu:2026nhi}.
Other realizations include singlet-doublet and inert-doublet dark matter,
models with new vector or spin-zero mediators, and photon-mediated
transitions~\cite{Borah:2026zwf,Wang:2026ytg,Zhu:2026dag,Du:2026lpa,
Yuan:2026djt,Leng:2026mwu,Wang:2026vpx,Yin:2026egz, Wang:2026ytg,Okada:2026upm, Lian:2026hpm, Gu:2026vto, Liang:2026coz}.
The candidate has also been studied in exothermic scenarios and in
halos containing both dark-matter mass
eigenstates~\cite{Dent:2026bji,deLima:2026shq,Baer:2026fpy,
An:2026pkc,Oliveira:2026jtn}.
Extending the nuclear-recoil analysis windows of XENONnT and
PandaX-4T to cover the LZ candidate energy would provide an independent
test of the high-recoil signal~\cite{XENON:2025vwd, PandaX:2024qfu, Dent:2026bji}.







In this work, we propose radiative $2\to3$ scattering,
$\chi N\to\chi N s$, as a mechanism for high-energy nuclear recoils.
The incoming and outgoing dark matter occupy the same ground state,
and the mass of the emitted invisible scalar $s$ sets the scattering
threshold. For a $1~\mathrm{TeV}$ dark-matter particle incident on xenon
at $700~\mathrm{km/s}$, the available center-of-mass kinetic energy is
about $300~\mathrm{keV}$. Scalar masses of $200$--$350~\mathrm{keV}$
therefore select the high-speed tail of the halo distribution and
suppress low-energy recoils. 
Earlier work studied a radiative $2\to3$ benchmark in which an
effectively massless scalar is emitted and $t$-channel exchange
of the same scalar enhances low-energy nuclear
recoils~\cite{Curtin:2013qsa, Curtin:2014afa}.

In our scenario, the minimum invariant mass $m_\chi+m_s$ of the
outgoing $\chi s$ pair sets the recoil endpoints at fixed incident
speed.
For a slowly varying squared
amplitude, three-body phase space shapes the recoil spectrum
through the continuum of accessible $\chi s$ invariant masses.
At either recoil endpoint, the $\chi s$ invariant mass takes its
minimum value.
Following the halo-independent approach~\cite{Fox:2010bu,Fox:2010bz},
we consider two recoil energies $E_L<E_H$ with the same minimum
incident speed $v_{\min}$ for a single isotope. At any incident
speed above $v_{\min}$, the larger momentum transfer at $E_H$
opens a wider range of $\chi s$ invariant masses. In the
heavy-dark-matter and constant-amplitude limits, the ratio of
the form-factor-corrected differential rates at $E_H$ and $E_L$
is about $(E_H/E_L)^{3/4}$ for any halo speed distribution
that populates both recoil energies. For inelastic $2\to2$
scattering with $\delta=m_s$, the recoil endpoints coincide
with those of the radiative process at each incident speed,
and the corresponding rate ratio is unity in the constant-amplitude
limit~\cite{Bozorgnia:2013hsa,Fox:2013pia}. Three-body phase space
thus favors the higher-energy member of an equal-$v_{\min}$ pair.

We illustrate this mechanism in a model with a light scalar and a
massive vector mediator. We calculate the recoil spectrum in natural
xenon by summing over its isotopes with their respective abundances,
kinematic thresholds, and nuclear form factors. For our benchmark,
we compare the radiative and inelastic spectra at matched thresholds
and equal integrated rates. The radiative spectrum is higher from
about $180$ to $580~\mathrm{keV}$, an interval that includes the LZ
candidate. We use the LZ exposure and nuclear-recoil detection
efficiency to determine the coupling product that gives one expected
signal event in a recoil-energy window around the candidate. At fixed
masses and halo distribution, this normalization also fixes the
expected number of radiative events below and above that window, providing a spectral test of the mechanism as more recoil data become available.
\\

\section{$2\to 3$ Scattering and Recoil Kinematics}
\begin{figure}[htb]
	\centering
	\includegraphics[width=0.45\linewidth]{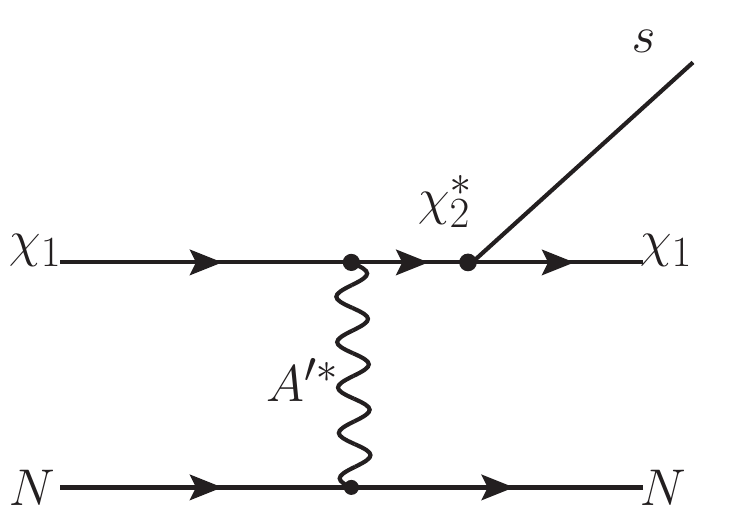} 
    \includegraphics[width=0.45\linewidth]{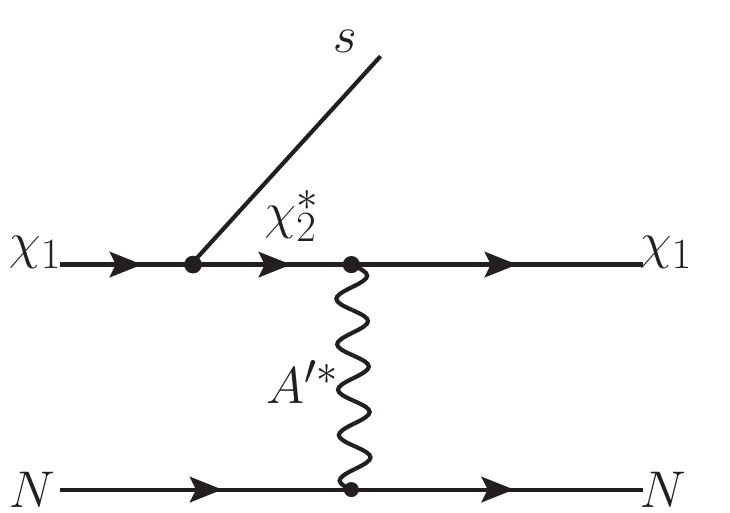}
	\caption{Tree-level diagrams for $\chi_1 N\to\chi_1 N + s$. The scalar $s$ is emitted from the dark matter line before or after $A'$ exchange.
        }
	\label{fig:2_to_3_feynmandiagrams}
\end{figure}

A dark-matter particle can produce an endothermic nuclear recoil
without changing its mass. We consider the $2\to 3$ process
$\chi_1 N\to\chi_1 N + s$, in which a real scalar $s$ is emitted
from the dark-matter line and the target nucleus $N$ remains in
its ground state. The model contains two Dirac fermions with
masses $m_{\chi_1}$ and $m_{\chi_2}=m_{\chi_1}+\Delta$, a massive
vector mediator $A'_\mu$, and a light scalar $s$. The relevant
interactions are
\begin{equation}
\begin{aligned}
\mathcal L_{\rm int}={}&
g_1 A'_\mu\bar\chi_2\gamma^\mu\chi_1
+g_2 s\bar\chi_2\chi_1+\mathrm{h.c.}\\
&+g_{q}A'_\mu\sum_f\bar q_f\gamma^\mu q_f ,
\end{aligned}
\label{eq:radiative-lagrangian}
\end{equation}
where $f$ runs over all flavor quarks and all couplings are taken
real. At leading order, the quark vector current gives equal proton and
neutron couplings, $g_p=g_n\equiv g_N=3g_q$~\cite{Boveia:2016mrp}. The vector exchange and the
scalar emission each change the dark-fermion state, so the two
tree-level diagrams, with emission before or after scattering,
both connect an incoming $\chi_1$ to an outgoing $\chi_1$, see Fig~\ref{fig:2_to_3_feynmandiagrams}.

The scalar mass $m_s$ sets the production threshold, whereas
$\Delta$ sets the threshold for producing an on-shell $\chi_2$.
For a given target (Xe, $m_{\chi_1} \simeq {\rm TeV}$), the hierarchy
\begin{equation}
m_s<\frac12\mu v_{\max}^2 \simeq 400{~\rm keV} <\Delta,
\quad \mu=\frac{m_{\chi_1} m_N}{m_{\chi_1}+m_N},
\label{eq:radiative-offshell-window}
\end{equation}
allows the massive scalar emission at sufficiently high speeds, 
but forbids
the inelastic scattering $\chi_1 N\to\chi_2 N$
throughout the halo speed range. 
Here $m_N$ is the nuclear mass and $v_{\max}$ is the
maximum incident speed of DM $\chi_1$ in the laboratory frame. 
If the intermediate $\chi_2$ state can go on shell and is narrow, 
the recoil spectrum reduces to that of conventional $2\to2$ inelastic scattering.
We focus on the off-shell regime.

To isolate the effect of the extra particle, we compare with
endothermic inelastic scattering $\psi_1 N\to\psi_2 N$ through the same
vector current, with masses $m_{\psi_1}$ and
$m_{\psi_2}=m_{\psi_1}+\delta$~\cite{Bramante:2016rdh}. Setting
$\delta=m_s$ matches the fixed mass of $\psi_2$ to the minimum
invariant mass of the invisible $\chi_1$-$s$ system. The two channels then have
identical kinematic recoil endpoints at a given incident speed,
including in exact kinematics. They differ in how they
populate the interval between those endpoints.

Let $q=\sqrt{2m_N E_R}$ be the momentum transfer at recoil
energy $E_R$, and write the invariant mass of the $\chi_1 \, s$ pair
as $W=m_{\chi_1}+m_s+\epsilon$, where $\epsilon$ is the internal kinematic energy of $\chi_1$-$s$ system. At fixed DM incident speed $v$, energy
conservation gives $\epsilon=vq\cos\theta-q^2/(2\mu)-m_s$ to leading order in the heavy-particle expansion, where $\theta$ is
the angle between the incoming dark matter and the nuclear recoil.
Putting $\theta = 0$, the maximum allowed $\epsilon$ is therefore
\begin{equation}
\begin{aligned}
D(E_R,v)&=q\,[v-v_{\min}(E_R)] =vq-\frac{q^2}{2\mu}-m_s,\\
v_{\min}(E_R)& = \frac{q^2/(2\mu) + m_s}{q}.
\end{aligned}
\label{eq:radiative-energy-budget}
\end{equation}
At $\epsilon=0$, $\chi_1$ and $s$ are at rest in their common
center-of-mass frame, so the pair has invariant mass $W=m_{\chi_1}+m_s$.
The nuclear recoil kinematics then coincide with those of
endothermic $2\to2$ scattering with $\delta=m_s$.
Both processes require $v\geq v_{\min}(E_R)$, or equivalently
$D\geq0$, but only the three-body process samples a continuum
of invariant masses, with $0\leq\epsilon\leq D$. 
Replacing $m_s$ with $\delta$ in $v_{\min}(E_R)$ gives the
corresponding expression for inelastic $2\to2$ scattering.

At fixed incident speed $v$, the condition $D \geq 0$ defines
the common recoil interval $E_R^-(v)\leq E_R\leq E_R^+(v)$
for the two processes with $\delta=m_s$. The endpoints are
\begin{equation}
E_R^\pm(v)=\frac{\mu^2}{2m_N}
\left(v\pm\sqrt{v^2-\frac{2m_s}{\mu}}\right)^2,
\label{eq:recoil-endpoints}
\end{equation}
provided $v\geq v_{\rm thr} \equiv \sqrt{2m_s/\mu}$. Below $v_{\rm thr}$, no recoil
is kinematically allowed in either channel.

For $m_{A'}^2\gg q^2$ and $\Delta \gg E_s$, where $E_s$ is
the scalar energy, both squared amplitudes are approximately
independent of the recoil variables after the common nuclear
response is factored out. For equal vector couplings in the
two models, the spin-averaged squared amplitudes reduce to
\begin{equation}
\overline{|\mathcal M_{2\to3}|^2}\simeq
\frac{4g_2^2}{\Delta^2}\, M^2_0, \qquad \overline{|\mathcal M_{2\to2}^{\rm inelastic}|^2}\simeq M_0^2,
\label{eq:radiative-amplitude-comparison}
\end{equation}
where $M^2_0$ is independent of recoil energy~\cite{Bramante:2016rdh}. The factor of four comes
from the constructive sum of the two radiation diagrams at the
amplitude level. 

\begin{figure}[htb]
    \centering
    \includegraphics[width=0.9\linewidth]{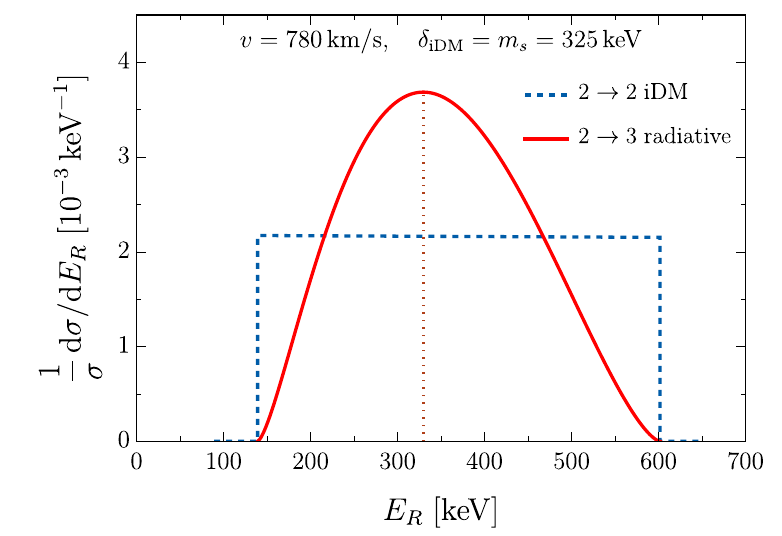}
\caption{Normalized differential recoil spectra
$(1/\sigma)\,d\sigma/dE_R$ for radiative $2\to3$ scattering
(solid) and inelastic $2\to2$ scattering (dashed) at fixed
incident speed $v=780~\mathrm{km/s}$ and matched thresholds
$\delta=m_s=325~\mathrm{keV}$. The vertical dotted line marks
$E_R^{\rm peak}=\mu^2v^2/(2m_N)$. The nuclear form factor is
set to unity.}
    \label{fig:fixed-speed-recoil-ratio}
\end{figure}

The additional phase space in the 3-body channel is
\begin{equation}
I(D)=\int_0^D d\epsilon\,\sqrt{\epsilon(\epsilon+2m_s)},
\label{eq:radiative-phase-space-kernel}
\end{equation}
with $I(D)=0$ for $D\leq0$. 
When $0 < D < m_s$, we have the expansion
\begin{align}
I(D)=\frac23\sqrt{2m_s}\,D^{3/2}
\left[1+\frac{3D}{20m_s}
+\mathcal O\!\left(\frac{D^2}{m_s^2}\right)\right].
\label{eq:ID-expansion}
\end{align}
In the heavy-$\chi_1$ limit, the
integrand is the magnitude of the scalar momentum in the
$\chi_1s$ rest frame. 
Within the common recoil interval, their ratio is
\begin{equation}
\frac{d\sigma_{2\to3}/dE_R}{d\sigma_{2\to2}/dE_R}
\simeq \frac{4g_2^2}{\Delta^2}\,
       \frac{I[D(E_R,v)]}{4\pi^2}.
\label{eq:radiative-differential-ratio}
\end{equation}
The squared-amplitude ratio sets the normalization, while
the three-body phase space supplies all the recoil-energy
dependence of this ratio.


At fixed $v>v_{\rm thr}$, the ratio in
Eq.~\eqref{eq:radiative-differential-ratio} vanishes as
$D^{3/2}$ toward either recoil endpoint and peaks at
$E_R^{\rm peak}=\mu^2v^2/(2m_N)$, where $D$ is maximal.
Fig.~\ref{fig:fixed-speed-recoil-ratio} shows the two
recoil spectra normalized to unit area. With the nuclear
form factor set to unity, the inelastic $2\to 2$ spectrum is flat.
The $2\to3$ spectrum lies above the flat spectrum at intermediate recoil energies and below it near both endpoints.
\\

\section{Features of the recoil spectra}
With $\delta=m_s$, the two processes have the same minimum
incident speed $v_{\min}$ at each recoil energy, but they weight the
incident speeds differently. For a single target isotope,
let $\widetilde R_i(E_R)\equiv F_N^{-2}(q)\,dR_i/dE_R$ be the
signal event rate spectrum after dividing out the common nuclear form
factor squared~\cite{Anand:2013yka}, 
where $i=2\to 2,2 \to 3$ labels the two channels. In the
constant-amplitude limit,
\begin{align}
\widetilde R_{\rm 2\to 2}(E_R)&\propto
\int_{v_{\min}(E_R)}^{v_{\max}}dv\,\frac{f(v)}{v},
\label{eq:equal-vmin-rate-two}\\
\widetilde R_{\rm 2\to3}(E_R)&\propto
\int_{v_{\min}(E_R)}^{v_{\max}}dv\,\frac{f(v)}{v}\,
I\bigl(q[v-v_{\min}(E_R)]\bigr), \nonumber
\label{eq:equal-vmin-rate-three}
\end{align}
where $f(v)$ is the normalized DM speed distribution in the
laboratory frame~\cite{Baxter:2021pqo}. The additional 3-body phase-space factor suppresses
the contribution from speeds just above $v_{\min}$.

The function $v_{\min}(E_R)$ has a minimum at
$E_c=\mu m_s/m_N$. Two recoil energies $E_L<E_c<E_H$ on
opposite sides of this minimum, have the same $v_{\min}$ if
\begin{equation}
E_L E_H=E_c^2 \quad \Rightarrow 
\quad   v_{\min}(E_L) = v_{\min}(E_H) .
\label{eq:equal-vmin-energy-pair}
\end{equation}
For the spin-independent two-body channel considered here, the
two energies probe the same halo integral and hence satisfy
$\widetilde R_{\rm 2\to2}(E_H)=\widetilde R_{\rm 2\to2}(E_L)$, as in the
inelastic shape test~\cite{Bozorgnia:2013hsa}.

For the three-body process, let $q_j=\sqrt{2m_NE_j}$ and
$D_j=q_j(v-v_{\min})$ for $j=L,H$. Inserting
Eq.~\eqref{eq:ID-expansion} into the speed integral and expanding
the rate ratio to next-to-leading order gives
\begin{equation}
\frac{\widetilde R_{2\to3}(E_H)}{\widetilde R_{2\to3}(E_L)}
\simeq
\left(\frac{E_H}{E_L}\right)^{3/4}
\left[1+\frac{3(q_H-q_L)}{20m_s}
\frac{J_{5/2}}{J_{3/2}}\right],
\end{equation}
where $J_n\equiv\int_{v_{\min}}^{v_{\max}}dv\,
[f(v)/v](v-v_{\min})^n$ is common to both recoil energies.

For $m=1~\mathrm{TeV}$ and $m_s=325~\mathrm{keV}$ on
$^{131}\mathrm{Xe}$, $E_L=240~\mathrm{keV}$ is paired with
$E_H\simeq349~\mathrm{keV}$. For a truncated Maxwellian halo with
$(|v_0|,v_{\rm esc},v_{\oplus}, v_\circledast)=(238,544,(29.2, -0.1, 5.9),(11.1, 12.2, 7.3))~\mathrm{km/s}$~\cite{Baxter:2021pqo},
the high-energy branch reaches $D_H\simeq62~\mathrm{keV}$ at
$v_{\max}\approx797.7~\mathrm{km/s}$. The leading-order rate ratio
is $1.324$, and the NLO term increases it by $0.26\%$.
Integrating the unexpanded kernel gives a ratio of $1.328$.
\\

\section{Xenon recoil spectra}
For natural xenon, we sum the recoil spectra of the individual isotopes $A$,
each with its own nuclear form factor and kinematic threshold. For
equal couplings to protons and neutrons, the differential rate per unit
detector mass in channel $i = 2\to2,~ 2\to 3$ is
\begin{align}
\frac{dR_i^{\rm Xe}}{dE_R}
&= C_i \sum_A \xi_A A^2 |F_A(q_A)|^2
   \int_{v_{\min,A}}^{v_{\max}}\!dv\,
   \frac{f(v)}{v}I_i(D_A),
\label{eq:xe-recoil-rate}
\end{align}
where $\xi_A$ is the isotope mass fraction~\cite{JRC58758}, $q_A=\sqrt{2m_AE_R}$,
and $C_i$ is independent of $E_R$ and $A$. With $\delta=m_s$, both
channels have $v_{\min,A}=m_s/q_A+q_A/(2\mu_A)$, where $\mu_A$ is the
DM--nucleus reduced mass for the isotope $A$. 
The kernels are $I_{2\to2}=1$ for inelastic $2\to2$ scattering and
$I_{2\to3}=I(D_A)$ for radiative $2\to3$ scattering, with
$D_A=q_A(v-v_{\min,A})$. Only isotopes with
$v_{\min,A}(E_R)<v_{\max}$ contribute at a given recoil energy.

\begin{figure}[bht]
    \centering
    \includegraphics[width=0.9\linewidth]{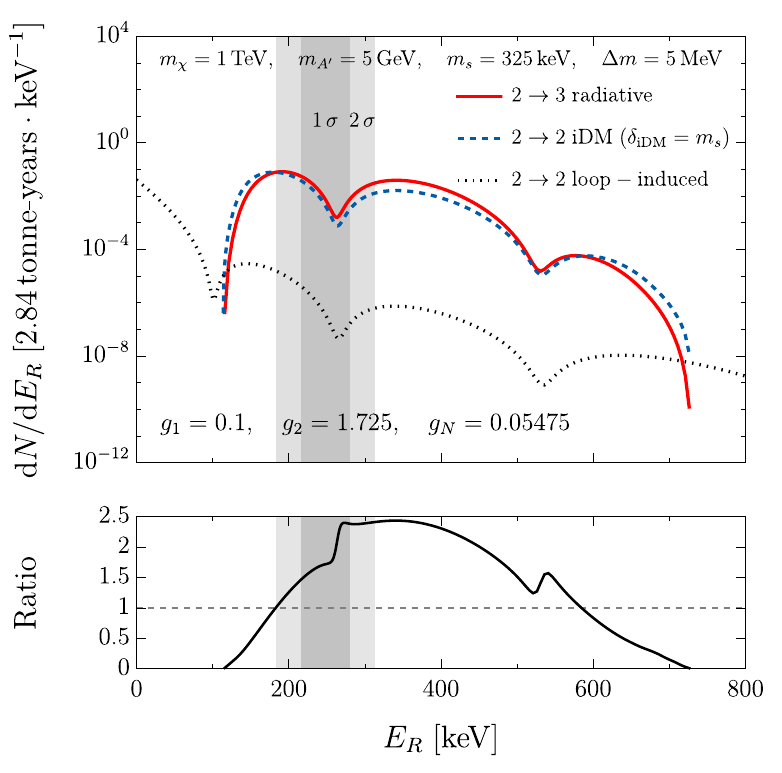}
    \caption{
Expected event spectra $dN/dE_R$ in natural xenon for radiative
$2\to3$ scattering (solid), inelastic $2\to2$ scattering (dashed), and
loop-induced elastic scattering (dotted).
We take $m_{\chi_1}=1~\mathrm{TeV}$, $m_{A'}=5~\mathrm{GeV}$, $\Delta=5~\mathrm{MeV}$ and match the
endothermic splitting to the emitted scalar mass,
$\delta=m_s=325~\mathrm{keV}$. 
The dark and light gray bands indicate the $1\sigma$ and $2\sigma$
recoil-energy intervals around the LZ candidate, respectively.
The loop-induced elastic $2\to 2$ spectrum 
retains its predicted normalization relative to the radiative signal.
The spectra are shown as functions of true nuclear recoil energy for an exposure of $2.84$ tonne-years, before applying the LZ
nuclear-recoil detection efficiency.
We normalize the inelastic $2 \to 2$ spectrum to the radiative $2\to 3$ spectrum. 
The lower panel shows the ratio of the $2\to 3$ spectrum to this $2\to 2$ 
inelastic spectrum.
}
    \label{fig:signal_spectrum1}
\end{figure}

Fig.~\ref{fig:signal_spectrum1} compares the two spectra for
$m_{\chi_1}=1~\mathrm{TeV}$ and $\delta=m_s=325~\mathrm{keV}$.
The form-factor minima occur at slightly different recoil energies for
different isotopes, leaving dips of finite depth in the summed spectra.
After normalization to the same integrated rate, the radiative $2\to 3$
spectrum lies above the inelastic $2\to2$ spectrum between about
$180$ and $580~\mathrm{keV}$ and below it near both recoil
endpoints. At the LZ candidate energy of $248~\mathrm{keV}$, the
ratio of the two spectra is about $1.6$. The suppression near
both endpoints persists after averaging over halo speeds and
summing over xenon isotopes, as the available three-body phase
space vanishes at each isotope's kinematic boundaries.
The dip near $260~\mathrm{keV}$ arises from the coherent xenon
density response in our vector-mediated benchmark. Interactions
that probe nuclear spin, such as those involving axial-vector~\cite{Elahi:2026vlm} or
pseudoscalar~\cite{Alhazmi:2026efz} quark currents, could weaken this dip 
without shifting the kinematic endpoints for a given isotope at fixed masses

Two-$A'$ box and crossed-box diagrams also induce elastic
$\chi_1 A\to\chi_1 A$ scattering, concentrated at low $E_R$
(dotted line in Fig.~\ref{fig:signal_spectrum1}). 
At fixed masses, the radiative
rate scales as $(g_1g_2g_{q})^2$, while the elastic rate scales as
$(g_1g_{q})^4$. A large $g_2$ therefore allows a small $g_1g_{q}$
at fixed radiative yield. Using the scalar-operator loop function
$F_3$ in \cite{Bell:2018zra,Duerr:2019dmv}, we estimate
$N_{\rm el}\simeq \{0.3,1.2\times 10^{-3}, 3.7 \times 10^{-5}, 4.8 \times 10^{-5}\}\times(g_1g_{N}/0.005)^4$ events
separately in $5$--$100~\mathrm{keV}$, $100$--$215.5~\mathrm{keV}$, $215.5$--$280.5~\mathrm{keV}$ and $300$--$650~\mathrm{keV}$ intervals, a negligible
contribution for the benchmark considered.
\\

\section{Viable parameter space and constraints}
The same coupling $g_2$ that allows a small loop-induced elastic rate
but large radiative rates also governs dark-matter annihilation into
$ss$ at freeze-out.

For two nearly degenerate Dirac species ($\Delta \ll m_{\chi_1}$) 
in chemical equilibrium
at freeze-out, we take $g_2\gg g_1$ and neglect the $A's$ and
$A'A'$ channels. In the light-scalar limit, the dominant
$\chi_i\bar\chi_i\to ss$ channel has p-wave
$\sigma v_{\rm rel}\simeq 3g_2^4v_{\rm rel}^2/(128\pi m_{\chi_1}^2)$
\cite{An:2016kie}. Accounting for both species and their
antiparticles, $\Omega_\chi h^2=0.12$ requires
\begin{equation}
g_2 \simeq 1.75
\left(\frac{m_{\chi_1}}{1~\mathrm{TeV}}\right)^{1/2}
\left(\frac{x_f}{26}\right)^{1/2}
\left(\frac{g_*}{90}\right)^{-1/8},
\label{eq:relic-g2}
\end{equation}
where $x_f=m_{\chi_1}/T_f$ and $g_*$ is evaluated at $T_f$. We compute the benchmark relic abundance with
micrOMEGAs~6.2.3~\cite{Alguero:2023zol} at tree level, neglecting
Sommerfeld enhancement from light-scalar exchange~\cite{An:2016kie}, which can reduce the relic-preferred value of $g_2$ but have little impact on our main result.

\begin{figure}[htb]
    \centering
    \includegraphics[width=0.9\linewidth]{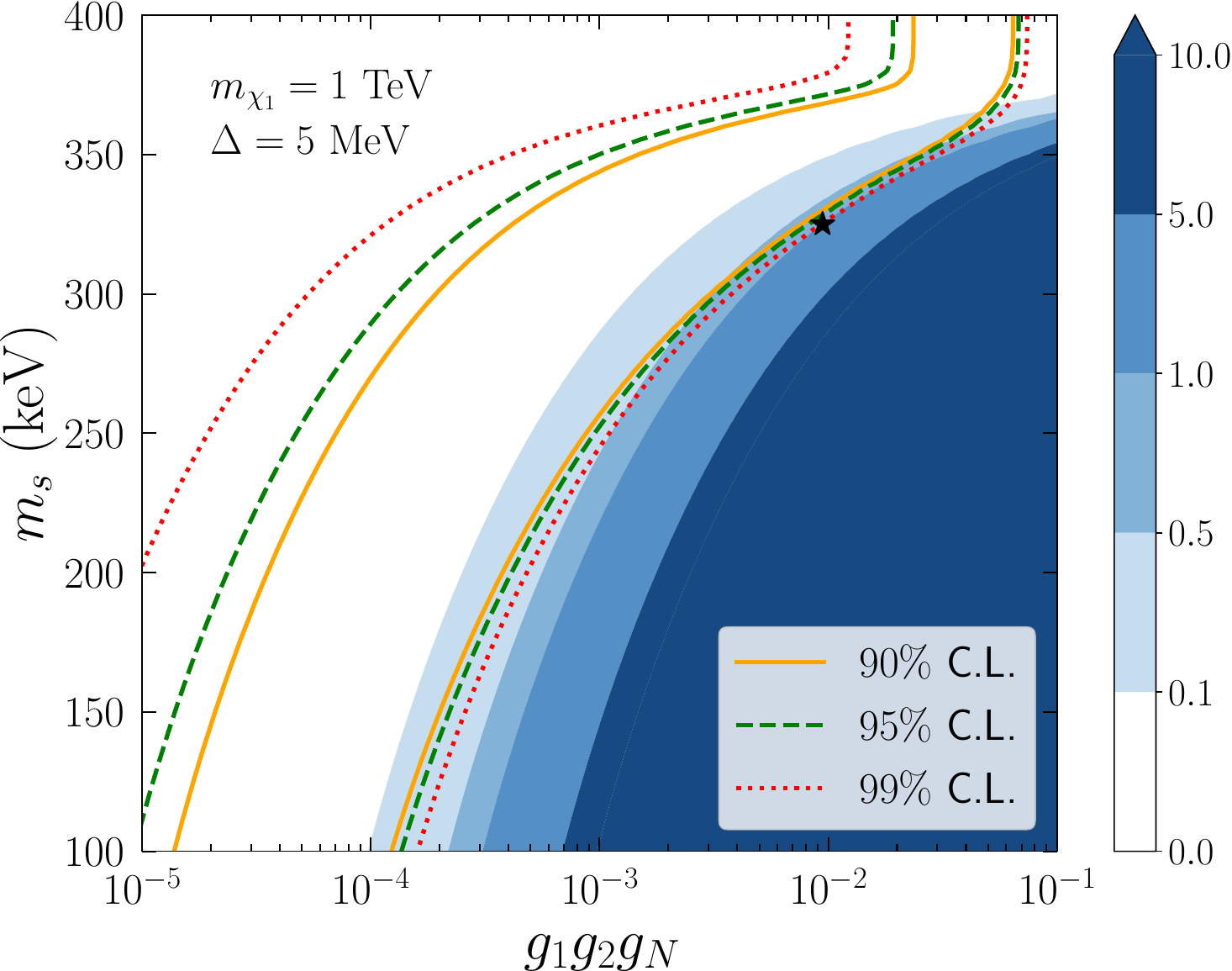}
    \includegraphics[width=0.9\linewidth]{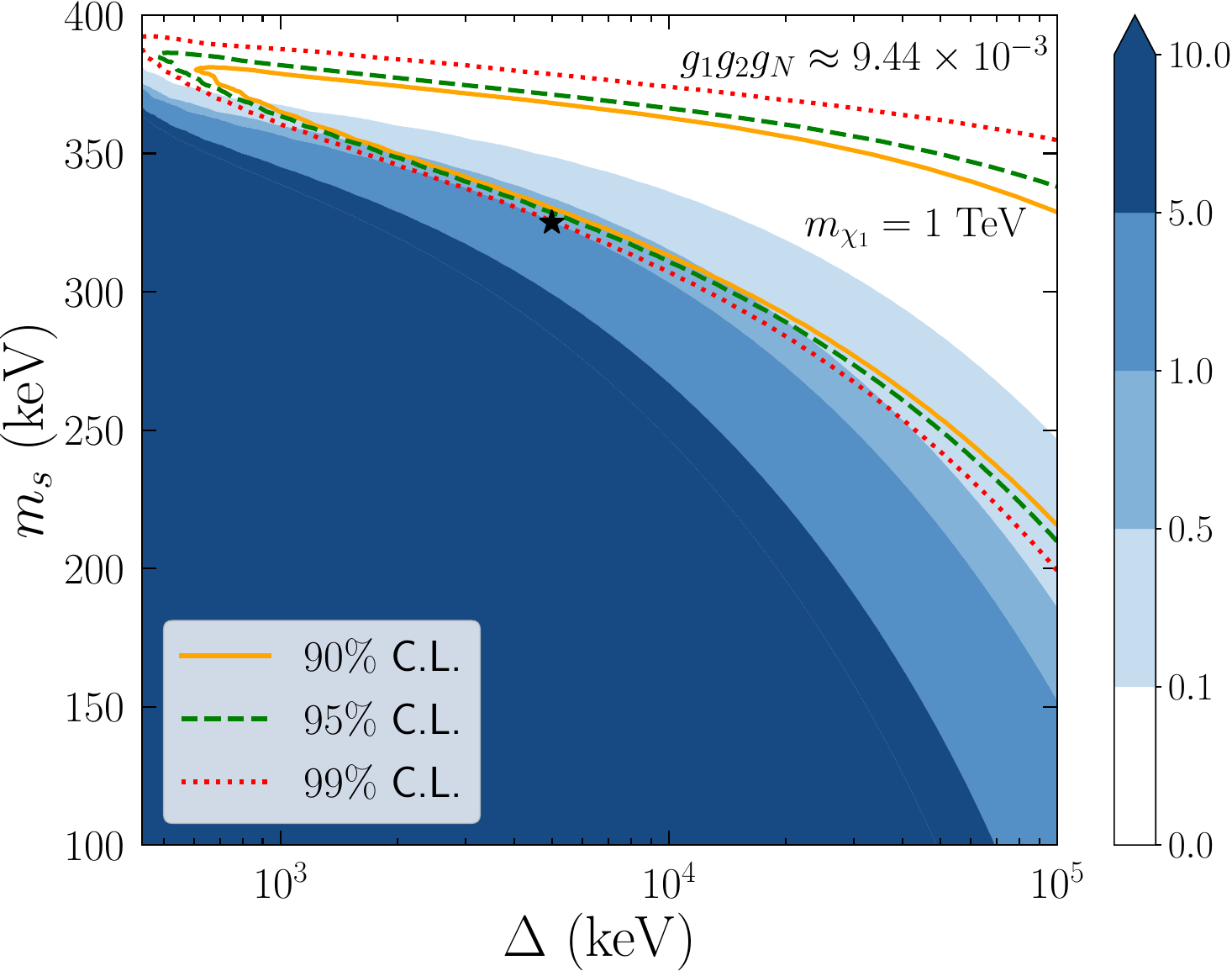}
    \caption{Fit to the LZ high-recoil candidate in the $(m_s, |g_1g_2g_N|)$ plane (upper panel) and the $(m_s, \Delta)$ plane (lower panel). The solid, dashed, and dotted curves
    correspond to fit-statistic levels of $90\%$, $95\%$,
    and $99\%$ confidence level, respectively. Blue shaded color shows the event number in $215.5$-$280.5~\mathrm{keV}$ interval. }
    \label{fig:ms_coupling_m_contour-2}
\end{figure}

With $g_2$ fixed by the relic abundance, we use the LZ exposure
and nuclear-recoil detection efficiency to determine
$g_1g_2g_{N}$ as a function of $m_s$ by requiring one expected
signal event in the recoil-energy window around the candidate.
We perform a $\chi^2$ analysis with three bins, where the signal events are taken to be $\{n_1, n_2, n_3\} = \{0, 1, 0\}$ separately for  $E_{nr} \in [100\ {\rm keV}, 215.5\ {\rm keV}]$, $ [215.5\ {\rm keV}, 280.5\ {\rm keV}]$, and $[300\ {\rm keV}, 650\ {\rm keV}]$, where the   data of last bin is given by Fig.~S4 of Ref.~\cite{LZ:2026axp}. 
We use the signal efficiency in \cite{LZ:2026axp} for $E_{nr} < 300$ keV region and employ a constant efficiency 0.95, for $E_{nr} \in [300\ {\rm keV}, 650\ {\rm keV}]$. Finally, a test statistic is performed by the log-likelihood ratio method.

For $m_{\chi_1}=1~\mathrm{TeV}$, and $m_{A'}=5~\mathrm{GeV}$,
Fig.~\ref{fig:ms_coupling_m_contour-2} shows the fit to the
LZ high-recoil candidate in the $(m_s,g_1g_2g_{N})$ and $(m_s,\Delta)$
plane. 
A heavier scalar raises the
minimum incident speed and reduces the available phase space,
requiring a larger coupling product or smaller mass gap $\Delta $. At
$m_s=325~\mathrm{keV}$ and $\Delta=5~\mathrm{MeV}$, one expected signal event in the
high-recoil selection requires
$g_1g_2g_{N}\simeq 9.44 \times 10^{-3}$.

An independent bound on $g_q$ comes from the ARGUS search
for nonelectromagnetic dijet decays of $\Upsilon(1S)$.
For a vector with universal quark couplings, virtual $A'$
exchange modifies this decay rate. Following
Ref.~\cite{Dobrescu:2021vak} and identifying $g_q=g_B/6$,
we obtain $|g_q|\lesssim 0.1166$ at
$m_{A'}=5~\mathrm{GeV}$, or equivalently
$g_q^2/m_{A'}^2\lesssim 5.44\times10^{-4}~\mathrm{GeV}^{-2}$
at this mass. The benchmark $(g_1,g_q)=(0.1,0.0183)$
reproduces the required LZ normalization and satisfies
this bound. 

We next consider the cosmological evolution of $A'$ and $s$. For the benchmark marked by the black star in the upper
panel of Fig.~\ref{fig:ms_coupling_m_contour-2}, decays and inverse decays $A'\leftrightarrow q\bar q$
keep the vector mediator in equilibrium with the SM bath~\cite{Ilten:2018crw}.
Although the light scalar $s$ has no direct coupling to the SM,
$A'A'\leftrightarrow ss$ and elastic $A's$ scattering initially keep it
in chemical and kinetic equilibrium with the bath. 
As the bath cools below $m_{A'}$, the $A'$ abundance becomes Boltzmann
suppressed, $s$ decouples while still relativistic at a temperature of $\mathcal{O}(1)$ GeV. 
Subsequent entropy release in the SM bath reduces the temperature
ratio to $T_s/T_\gamma\simeq0.50$ at $T_\gamma=10\,\mathrm{MeV}$.

We allow $s$ to decay into two effectively massless sterile
neutrinos, $s\to N_sN_s$. For $y_\nu=3\times10^{-9}$,
most decays occur after the $A'$-$s$ decoupling. Solving the coupled
Boltzmann equations for the number and energy densities of $s$ and $N_s$, we find $\Delta N_{\rm eff}\simeq0.041$ (see Appendix~\ref{app:early_universe}),
consistent with CMB constraints~\cite{Planck:2018vyg, AtacamaCosmologyTelescope:2025nti}.
\\

\section{Conclusion}
We have proposed radiative $2\to3$ scattering as a mechanism for
high-energy nuclear recoils from halo dark matter. In
$\chi N\to\chi N s$, the energy required to produce a massive $s$ sets the
scattering threshold while the incoming and outgoing dark matter
occupy the same ground state. When $\delta=m_s$, the recoil
endpoints coincide with those of inelastic $2\to2$ scattering at
every allowed incident speed. 
The recoil spectrum can nevertheless distinguish the two processes. 
In our benchmark, the ratio of the radiative $2\to3$ to inelastic
$2\to2$ event spectra exceeds unity from about $180$ to
$580~\mathrm{keV}$ and falls below unity near both recoil
endpoints. 
Three-body phase space determines the allowed recoil interval
for radiative $2\to3$ scattering and governs the spectral shape
within that interval when the squared amplitude is nearly constant.
The same kinematic result also applies to a local nucleon-level interaction such as $s(\bar\chi\chi)(\bar{N}N)/\Lambda^3$, where $N$ denotes a nucleon.

Our benchmark with $m_\chi=1~\mathrm{TeV}$, $m_s=325~\mathrm{keV}$,
and $m_{A'}=5~\mathrm{GeV}$ reproduces the thermal relic abundance
and satisfies the bounds on $A'$ and $s$ considered here. For the
LZ exposure of $2.84$ tonne-years, it predicts approximately $\{0,4.14,1, 3.73\}$ recoil events 
in $5$--$100~\mathrm{keV}$, $100$--$215.5~\mathrm{keV}$, $215.5$--$280.5~\mathrm{keV}$ and $300$--$650~\mathrm{keV}$ intervals, after applying the detection
efficiency.
Loop-induced elastic scattering contributes only about 0.43 events between $5$--$100~\mathrm{keV}$. If further events accumulate near
$248~\mathrm{keV}$, xenon experiments can test this benchmark by
searching for the accompanying signal at $100$--$200~\mathrm{keV}$ and $300$--$650~\mathrm{keV}$.
With sufficient exposure, the relative yields in the two windows
can also distinguish the radiative $2\to 3$ and inelastic $2\to 2$ benchmarks. 

\vspace{12pt}
\begin{acknowledgments}
\textit{Acknowledgements}.
R.D.F. and J.L. thank Yongheng Xu for helpful discussions on the experimental details of the LZ high-recoil search.
The work of J.L. is supported by the National Science Foundation of China under Grant No.~12235001, No.~12475103 and State Key Laboratory of Nuclear Physics and Technology under Grant No. NPT2025ZX11.
The work of X.P.W. is supported by the National Science Foundation of China under Grant No. 12375095, and the Fundamental Research Funds for the Central Universities. 
The authors gratefully acknowledge the valuable discussions and insights provided by the members of the Collaboration of Precision Testing and New Physics.
\end{acknowledgments}

\clearpage

\appendix

\section{Cosmological evolution}
\label{app:early_universe}

We describe the cosmological evolution of the particles and compute the resulting $\Delta N_{\rm eff}$ by solving their coupled Boltzmann equations.

\subsection{Post-freeze-out evolution}

We follow the light-sector evolution from the heavy-particle
freeze-out temperature $T_f$ through scalar decay. Sterile
neutrinos are denoted by $N_s$, distinct from nucleons.
We assume that $A'$ remains in kinetic and chemical equilibrium
with the SM because of the high decay rate~\cite{Ilten:2018crw}, so $T_{A'}=T_\gamma\equiv T$, and use Maxwell--Boltzmann
(MB) distributions
\[
f_{A'}=e^{-E_{A'}/T},
\quad
f_s=z_s e^{-E_s/T_s},
\quad
z_s=\frac{n_s}{n_s^{\rm eq}(T_s)}.
\]
The equilibrium density and mean energy are
\cite{EscuderoAbenza:2020cmq,Cirelli:2024ssz}
\begin{align}
n_i^{\rm eq}(T_i)
&=\frac{g_i m_i^2T_i}{2\pi^2}K_2(\frac{m_i}{T_i}),
\quad g_{A'}=3,\quad g_s=1,
\\
u_i(T_i)
&=3T_i+m_i\frac{K_1(\frac{m_i}{T_i})}{K_2(\frac{m_i}{T_i})}.
\end{align}
Thus $\rho_s=n_su_s(T_s)$ and $P_s=n_sT_s$, including
out-of-equilibrium abundances within this closure.
We initialize $T_s=T_f$ and $n_s=n_s^{\rm eq}(T_f)$.

The heavy-particle $\chi_{1,2}, \bar\chi_{1,2}$ history is imported from
micrOMEGAs 6.2.3~\cite{Alguero:2023zol}:
\begin{align}
n_{\rm heavy}
=&\sum_{i=1,2}(n_{\chi_i}+n_{\bar\chi_i})
=s_{\rm SM}Y_{\rm heavy},\\
\rho_{\rm heavy}\simeq & \,\,m_{\chi_1}n_{\rm heavy}.
\end{align}
Here we neglect its kinetic energy, the small mass-splitting
correction, and feedback on the imported freeze-out history.
After freeze-out is complete, $n_{\rm heavy}\propto a^{-3}$.

While the SM species share a common temperature, the bath
thermodynamic functions and expansion rate are
\begin{align}
\rho_{\rm bath}
&=\rho_{\rm SM}+n_{A'}^{\rm eq}u_{A'},
\\
P_{\rm bath}
&=Ts_{\rm SM}-\rho_{\rm SM}+n_{A'}^{\rm eq}T,
\\
H^2
&=\frac{8\pi}{3M_{\rm Pl}^2}
\left(\rho_{\rm bath}+\rho_s+\rho_{N_s}+\rho_{\rm heavy}\right).
\end{align}
Here $M_{\rm Pl}$ is the unreduced Planck mass, and the SM
thermodynamic functions exclude dark-sector contributions.
SM chemical potentials are neglected.

For the post-freeze-out scalar evolution, we retain
$A'A'\leftrightarrow ss$ and $A's\leftrightarrow A's$, together
with scalar decay and inverse decay into sterile neutrinos.
The sources unrelated to sterile decay are therefore
\begin{align}
C_0 &= C_{\rm pair},\\
Q_0 &= Q_{\rm pair}+Q_{\rm elastic}.
\end{align}
Here $Q_0$ is the net energy transferred from the thermal
bath to the scalar population.

We neglect heavy-catalyzed conversion $A'h\leftrightarrow sh$,
heat exchange between the scalar population and the SM bath
mediated by residual heavy particles, and the net scalar
source from $\chi_2$ decays and inverse decays, including
excited-state regeneration through heavy-particle collisions
and scattering with the SM bath.
Here $h=\chi_{1,2},\bar\chi_{1,2}$.
While the final numerical result $\Delta N_{\rm eff}$ for the benchmark considered here does not show much difference, we therefore neglect it in the calculation.

Including the decay terms defined below,
$C_s=C_0+C_{\rm dec}$ and $Q_s=Q_0+Q_{\rm dec}$,
the scalar moment equations are~\cite{EscuderoAbenza:2020cmq}
\begin{align}
\dot n_s+3Hn_s&=C_s,
\\
\dot\rho_s+3H(\rho_s+P_s)&=Q_s.
\end{align}
With $c_{V,s}={\rm d}u_s/{\rm d}T_s$, the temperature equations
are
\begin{align}
\dot T_s
&=\frac{Q_s-u_sC_s-3Hn_sT_s}{n_sc_{V,s}},
\\
\dot T
&=-\frac{3H(\rho_{\rm bath}+P_{\rm bath})+Q_0}
{{\rm d}\rho_{\rm bath}/{\rm d}T}.
\end{align}
Positive sources add scalar number or energy. The bath
loses only $Q_0$, not the internal exchange $Q_{\rm dec}$.
Residual heavy-annihilation heating is neglected.

For $A'A'\leftrightarrow ss$, we define the equilibrium event
density~\cite{Cirelli:2024ssz}
\begin{align}
\gamma_{A'A'\to ss}^{\rm eq}(T)
=&\frac12[n_{A'}^{\rm eq}(T)]^2
\langle\sigma v_{\rm Mol}\rangle
\nonumber\\
=&\frac{9T}{64\pi^4}
\int_{s_{\min}}^\infty {\rm d}s\,
s^{3/2}\lambda(s,m_{A'}^2,m_{A'}^2)\nonumber\\
& \times\sigma_{A'A'\to ss}(s)K_1(\frac{\sqrt{s}}{T}),
\end{align}
where $\lambda(s,m_1^2,m_2^2)
=\left[1-\frac{(m_1+m_2)^2}{s}\right]
 \left[1-\frac{(m_1-m_2)^2}{s}\right]$, and $s_{\min}=\max(4m_{A'}^2,4m_s^2)=4m_{A'}^2$ for our benchmark. 
The corresponding total-energy moment is
\cite{Fernandez:2021iti}
\begin{align*}
\mathcal E_{A'A'\to ss}^{\rm eq}(T)
=&\frac{9T}{64\pi^4}
\int_{s_{\min}}^\infty {\rm d}s\,
s^2\lambda(s,m_{A'}^2,m_{A'}^2)\\
& \times\sigma_{A'A'\to ss}(s)K_2(\frac{\sqrt{s}}{T}).
\end{align*}
Detailed balance then gives
\begin{align}
C_{\rm pair}
&=2\left[\gamma_{A'A'\to ss}^{\rm eq}(T)
-z_s^2\gamma_{A'A'\to ss}^{\rm eq}(T_s)\right],
\\
Q_{\rm pair}
&=\mathcal E_{A'A'\to ss}^{\rm eq}(T)
-z_s^2\mathcal E_{A'A'\to ss}^{\rm eq}(T_s).
\end{align}
No additional factor of two enters $Q_{\rm pair}$, since
$\mathcal E^{\rm eq}$ weights the total energy per reaction.
For temperature-independent microscopic inputs, the moments
needed for the relaxation analysis follow from
\[
\mathcal E^{\rm eq}
=T^2\frac{{\rm d}\gamma^{\rm eq}}{{\rm d}T},
\quad
\mathcal E_2^{\rm eq}
=T^2\frac{{\rm d}\mathcal E^{\rm eq}}{{\rm d}T}.
\]

For $A'(p_1)s(p_2)\to A'(p_3)s(p_4)$, starred quantities
refer to the center-of-mass frame, while unstarred energies
refer to the cosmological frame. The scalar energy and
momentum in the former are
\[
E_2^*=E_4^*=\frac{s+m_s^2-m_{A'}^2}{2\sqrt{s}},
\quad
p_*=\frac{\sqrt{s\lambda(s,m_{A'}^2,m_s^2)}}{2}.
\]
With boost factor $\gamma_B=(E_1+E_2)/\sqrt{s}$, averaging
over the scattering azimuth gives
\[
\overline{E_4-E_2}
=(\gamma_BE_2^*-E_2)(1-\cos\theta_*).
\]
This motivates the momentum-transfer cross section
\[
\sigma_T(s)
=\int {\rm d}\Omega_*\,(1-\cos\theta_*)
\frac{{\rm d}\sigma_{A's\to A's}}{{\rm d}\Omega_*}.
\]
After integration over the incoming orientations,
\[
Q_{\rm elastic}
=\frac{3z_s}{32\pi^4}
\int_{(m_{A'}+m_s)^2}^\infty {\rm d}s\,
s\sqrt{\lambda(s,m_{A'}^2,m_s^2)}\,
\sigma_T(s)\mathcal I_E(s),
\]
where
\begin{align*}
\mathcal I_E(s)
= &\int_{\mathcal D(s)}{\rm d}E_1\,{\rm d}E_2\,
e^{-E_1/T-E_2/T_s}\\
&\times \left[(E_1+E_2)\frac{s+m_s^2-m_{A'}^2}{2s}-E_2\right].
\end{align*}
Writing $E=E_1+E_2$ and $P=\sqrt{E^2-s}$, the integration
domain is
\[
E\ge\sqrt{s},\quad aE-bP\le E_2\le aE+bP,
\]
\[
a=\frac{s+m_s^2-m_{A'}^2}{2s},\quad
b=\frac{\sqrt{\lambda(s,m_{A'}^2,m_s^2)}}{2}.
\]
Performing the remaining energy integrals in the
unequal-temperature MB formalism~\cite{BeauchesneKats2021}
yields
\[
\mathcal I_E(s)
=\frac{s^2\lambda^{3/2}(s,m_{A'}^2,m_s^2)}{4}
\left(\frac1{T_s}-\frac1T\right)\frac{K_2(w)}{w^2},
\]
\[
w^2=\frac{m_{A'}^2}{T^2}+\frac{m_s^2}{T_s^2}
+\frac{s-m_{A'}^2-m_s^2}{TT_s}.
\]
Hence
\begin{align}
Q_{\rm elastic}
={}&\frac{3z_s}{128\pi^4}
\left(\frac1{T_s}-\frac1T\right)
\nonumber\\
&\times\int_{(m_{A'}+m_s)^2}^\infty {\rm d}s\,
s^3\lambda^2(s,m_{A'}^2,m_s^2)
\sigma_T(s)\frac{K_2(w)}{w^2}.
\end{align}
At a fixed bath temperature, we linearize the scalar number and energy
equations about $T_s=T$ and $z_s=1$ and compare the smaller relaxation
rate with the Hubble rate $H$.
For our benchmark, the slower relaxation rate equals $H$ at
$T\simeq 5.0\,\mathrm{GeV}$,
which we use as an estimate of the scalar decoupling temperature.
\subsection{Scalar decay and dark radiation}

We introduce one effectively massless Majorana sterile
neutrino through
\[
\mathcal L_{\rm dec}
=-\frac{y_\nu}{2}s\overline{N_s}N_s,
\quad
\Gamma_s=\frac{y_\nu^2m_s}{16\pi},
\]
where $\Gamma_s$ is the rest-frame width. The sterile
population is described by an MB moment closure,
\[
f_{N_s}=z_{N_s}e^{-p/T_{N_s}},
\quad
n_{N_s}^{\rm eq}=\frac{2T_{N_s}^3}{\pi^2},
\]
\[
z_{N_s}=\frac{n_{N_s}}{n_{N_s}^{\rm eq}},
\quad
\rho_{N_s}=3n_{N_s}T_{N_s}.
\]
Here, chemical equilibrium is not imposed.

The decay and inverse-decay sources into $s$ are
\cite{EscuderoAbenza:2020cmq}
\begin{align}
C_{\rm dec}
&=\frac{\Gamma_s m_s^2}{2\pi^2}
\left[z_{N_s}^2T_{N_s}K_1(\frac{m_s}{T_{N_s}})
-z_sT_sK_1(\frac{m_s}{T_s})\right],
\label{eq:scalar-decay-number}
\\
Q_{\rm dec}
&=\frac{\Gamma_s m_s^3}{2\pi^2}
\left[z_{N_s}^2T_{N_s}K_2(\frac{m_s}{T_{N_s}})
-z_sT_sK_2(\frac{m_s}{T_s})\right].
\label{eq:scalar-decay-energy}
\end{align}
The inverse terms vanish for an empty sterile population.
Without inverse decays, $Q_{\rm dec}=-m_s\Gamma_s n_s$,
since the energy weight cancels the decay time-dilation
factor. 
The sterile moments obey
\begin{align}
\dot n_{N_s}+3Hn_{N_s}&=-2C_{\rm dec},\\
\dot\rho_{N_s}+4H\rho_{N_s}&=-Q_{\rm dec}.
\label{eq:sterile-energy-evolution}
\end{align}
We initialize $n_{N_s}=\rho_{N_s}=0$ at $T_f$ and evolve
both moments continuously through scalar decay.
In the MeV epoch, $A'$ is Boltzmann suppressed and we
neglect $C_0$ and $Q_0$. The decay sources cancel in the
combined energy equation, giving
\begin{equation}
\frac{\mathrm d}{\mathrm d\ln a}
\left[a^4(\rho_s+\rho_{N_s})\right]
=a^4(\rho_s-3P_s).
\end{equation}
A nonrelativistic scalar can therefore enhance the
comoving energy ultimately carried by the sterile neutrinos.

The electromagnetic bath contains photons and
Fermi--Dirac $e^\pm$. We approximate active-neutrino
decoupling as instantaneous at $T_\gamma=2\,{\rm MeV}$,
with $\rho_{\nu_a}=7\pi^2T_\nu^4/40$ for three species.
Before decoupling, $T_\nu=T_\gamma$ and active neutrinos
are included in the common thermal bath.
Afterwards, $aT_\nu$ is constant, and the photon temperature
follows the bath energy equation with only the
electromagnetic component. The active-neutrino energy
density continues to contribute to $H$.

After scalar decay and $e^+e^-$ annihilation are complete,
we evaluate
\begin{equation}
\Delta N_{\rm eff}
=\frac{\rho_{N_s}}
{\frac78(\frac4{11})^{4/3}\rho_\gamma},
\qquad
\rho_\gamma=\frac{\pi^2}{15}T_\gamma^4.
\label{eq:sterile-delta-neff}
\end{equation}
We quote this contribution separately or use
$N_{\rm eff}\simeq3.045+\Delta N_{\rm eff}$ with the
fixed SM reference value of
Ref.~\cite{EscuderoAbenza:2020cmq},
without recalculating active-neutrino reheating corrections.

\bibliography{ref.bib}{}
\bibliographystyle{utphys28mod}

\end{document}